\documentclass[
    ,final            % use final for the camera ready runs
  ]
  {aipproc}

\layoutstyle{6x9}

\usepackage{epsfig}
\usepackage{graphicx}
\usepackage{psfrag}

\newcommand{\prd}[3]{Phys. Rev. \textbf{D{#1}} ({#2}) {#3}}

\begin{document}

\title{The Infrared Landau Gauge Gluon Propagator from Lattice QCD}

\author{O. Oliveira}{
  address={Centro de Física Computacional, Departamento de Física, Universidade de Coimbra, 3004-516 Coimbra, Portugal},
  email={orlando@teor.fis.uc.pt}
}

\author{P. J. Silva}{
  address={Centro de Física Computacional, Departamento de Física, Universidade de Coimbra, 3004-516 Coimbra, Portugal},
  email={psilva@teor.fis.uc.pt}
}

\begin{abstract}
The quenched Landau gauge gluon propagator is investigated in lattice QCD
with large assimetric lattices, accessing momenta as low as $q \sim 100$ MeV
or smaller. Our investigation focus on the IR limit of the gluon dressing 
function, testing the compatibility with recent solutions of the 
Dyson-Schwinger equations. In particular, the low energy parameters $\kappa$ 
and $\alpha (0)$ are measured.
\end{abstract}

\maketitle

%%%%%%%%%%%%%%%%%%%%%%%%%%%%%%%%%%%%%%%%%%%%
%% MAINMATTER
%%%%%%%%%%%%%%%%%%%%%%%%%%%%%%%%%%%%%%%%%%%%

%\section{Introduction and motivation}

The confinement of quarks in hadrons and the chiral symmetry breaking 
mechanism are believed to be linked to the low 
energy properties of QCD. In particular, the gluon and ghost propagators can 
provide us information on the mechanism of confinement. 

Two first principles approaches to non-perturbative problems are 
Dyson-Schwinger equations (DSE) and lattice methods. In lattice QCD, the gluon
propagator has been revisited a number of times. However, due to the lattice 
sizes used in previous studies, the access to the IR region was limited. 
For the Dyson-Schwinger solution for the gluon propagator, recently in
\cite{Lerche02,alkofer03} 
the low energy behaviour of the propagator was solved 
analytically. Moreover, the authors found a parametrization for the gluon 
dressing function that fitted the numerical solution of the DSE. The lattice 
being complementary to DSE, allows a check of compatibility between 
the two methods. In particular, we would like to check for the behaviour of 
small and zero momenta gluon propagator.

Our investigation uses pure gauge, Wilson action, SU(3) configurations, 
$\beta = 6.0$ ($a^{-1} = 1.885$ GeV), 
on large assimetric lattices\footnote{All configurations were generated with
MILC code \url{http://physics.indiana.edu/~sg/milc.html}.}: $16^3 \times 128$ 
and $16^3 \times 256$. They were generated using combinations of 
over-relaxation (OVR) and Cabibbo-Mariani (HB) updates. For the smaller 
(larger) lattice, a combined sweep of 7 OVR and 2 HB (7 OVR and 4 HB) was used
and 3000 combined sweeps for thermalization. The 160 (70) configurations were
saved with a separation of 3000 (1500) combined sweeps. 
Note that, due to the large extension on the time direction, the lowest 
momenta considered is about 93 MeV for the smallest lattice and 46 MeV for 
the largest lattice.

\begin{figure}[t]
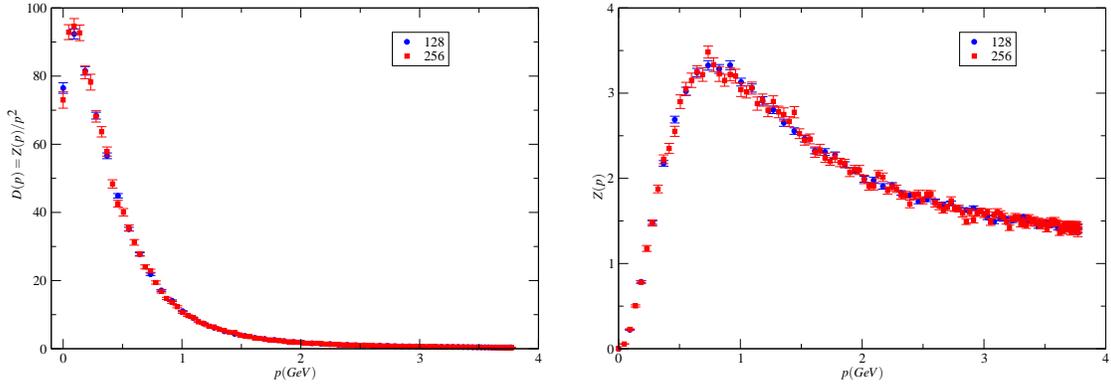

   \hspace*{+0.1cm}
   \begin{minipage}[b]{0.55\textwidth}
      \psfrag{EIXOX}{\begin{tiny}$p(GeV)$\end{tiny}}
      \psfrag{EIXOY}{\begin{tiny}$D(p)=Z(p)/p^2$\end{tiny}}
      \centering
      \includegraphics[origin=c,scale=0.30]{raw.eps}
   \end{minipage}
   \hfill
   \hspace*{-0.7cm}
   \begin{minipage}[b]{0.55\textwidth}
      \psfrag{EIXOX}{\begin{tiny}$p(GeV)$\end{tiny}}
      \psfrag{EIXOY}{\begin{tiny}$Z(p)$\end{tiny}}
      \centering
      \includegraphics[origin=c,scale=0.30]{temporal128256.eps}
   \end{minipage}
\caption{Gluon propagator. Statistical errors were computed using the
jackknife procedure.} \label{prop}
\end{figure}

For notation and details see \cite{gribovgluon}. The data for the Landau 
gauge gluon propagator\footnote{For gauge fixing we used the
steepest descent Fourier accelerated method described in \cite{davies}.}
\begin{equation}
  \langle A^a_\mu (p) ~  A^a_\mu (p') \rangle ~ = ~ V \, 
   \delta \left( p + p' \right) \, \delta^{a b}
   \left( \delta_{\mu\nu} - \frac{p_\mu p_\nu}{p^2} \right) \,
   \frac{ Z( p^2 ) }{p^2} 
\end{equation}
shows finite volume effects. To disentangle the finite volume effects, our 
data was compared with the results reported in \cite{leinweber}. No essential 
differences were seen between our pure temporal data and the Leinweber 
\textit{et al} data. From now on, we refer only to pure temporal data -- see
figure \ref{prop}. Note that the data seems to support a vanishing zero 
momenta gluon propagator $D(p^2)$.

To test the compatibility between lattice and the DSE results 
\cite{alkofer03}, our data for the gluon dressing function was fitted to
\begin{equation}
 Z(p^2) = \omega
          \left(\frac{p^2}{\Lambda_{QCD}^2+p^2}\right)^{2\kappa}
          \left(\alpha(p^2)\right)^{-\gamma} \, , \label{Zp2}
\end{equation}
where $\gamma = -13/22$, 
using two ansatz for the running coupling \cite{alkofer03,fischerexpr37}:
\begin{eqnarray}
  \alpha_1 (p^2) &  =  &
       \frac{1}{1+\frac{p^2}{\Lambda_{QCD}^2}}
       \bigg[\alpha(0) ~ + ~ 
             \frac{p^2}{\Lambda_{QCD}^2}
                 \frac{4\pi}{\beta_0}
                 \Big( \frac{1}{ln(p^2/\Lambda_{QCD}^2)} ~ - ~
                       \frac{1}{p^2/\Lambda_{QCD}^2-1}\Big)\bigg] \, , \\
  \alpha_2(p^2) &  =  &
   \frac{\alpha(0)}{ln\left[e \, + \, a_1(p^2/\Lambda_{QCD}^2)^{a_2}\right]}
   \, ,
 \label{alfa}
\end{eqnarray}
with $\beta_0 = 11$.

The IR propagator was investigated fitting the lattice data to 
\begin{equation}
Z_{IR1}(p^2)=\omega\left(p^2\right)^{2\kappa}\, \hspace{0.7cm}
Z_{IR2}(p^2)=\omega\left(\frac{p^2}{\Lambda_{QCD}^2+p^2}\right)^{2\kappa}
\label{eqir}
\end{equation}
Only when the first three lowest momenta of the larger lattice 
($\vert q \vert \leq 139 MeV$) were considered, we were able to fit the IR 
analytical solution of DSE $Z_{IR1}$ with an acceptable 
$\chi^2/d.o.f. = 0.39$,  meaning that the solution 
$Z(q^2)\sim(q^{2})^{2\kappa}$ seems to be valid only for momenta below 150 MeV.
The measured\footnote{Statistical errors for fit parameters were computed with
2000 bootstrap samples.} $\kappa=0.5003^{+32}_{-29}$ does not provide a clear
answer about the zero momenta gluon propagator.
The results for the IR2 approximation are shown in table \ref{ir2}. For both 
lattices, the fit to the largest range of momenta supports an infrared finite
gluon propagator.

\begin{table}[h]
\begin{tabular}{lccccc}
\hline
\tablehead{1}{c}{b}{Lattice} & 
\tablehead{1}{c}{b}{Range} & \tablehead{1}{c}{b}{$\kappa$}
  & \tablehead{1}{c}{b}{$\Lambda_{QCD}$ (MeV)}
  & \tablehead{1}{c}{b}{$\chi^2 /d.o.f.$}  \\
\hline
$16^3\times 128$ & $\|q\|\leq 461 MeV$ & $0.5020^{+46}_{-49}$ & $429^{+10}_{-9}$&  $0.50$ \\
$16^3\times 128$ & $\|q\|\leq 553 MeV$ & $0.5122^{+42}_{-46}$ & $403^{+8}_{-7}$ &  $1.41$ \\
$16^3\times 256$ & $\|q\|\leq 644 MeV$ & $0.5199^{+26}_{-31}$ & $377^{+5}_{-5}$ &  $1.19$ \\
\hline
\end{tabular}
\caption{$Z_{IR2}$ fits.}
\label{ir2}
\end{table}

The fits to all data distinguishes the functional forms $\alpha_1 (p^2)$ and
$\alpha_2 ( p^2 )$ for the running coupling constant. For $\alpha_1 ( p^2 )$, 
the $\chi^2/d.o.f. > 2$. Moreover, $\alpha (0) \sim 10$ is much larger than
the DSE result: $\alpha (0) = 2.972$ \cite{alkofer03,Lerche02}. However,
if one uses $\alpha_2 ( p^2 )$, the lattice data is 
well described by (\ref{Zp2}) - see table \ref{dse2}. 
$\alpha (0)$ can be computed from the asymptotic behaviour of the QCD $\beta$ 
function and $\alpha_2 (p^2)$. Then,
$ \alpha (0) = (4\pi/\beta_0)a_2$ giving $\alpha (0) =2.78^{+2}_{-2}$  and
$2.74^{+1}_{-2}$ for the smaller and larger lattices, respectively.
The analysis of all momenta favours a finite zero momenta gluon propagator.  
Our figures for $\kappa$ are smaller than those reported in the DSE studies,
$\kappa \sim 0.595$ \cite{Lerche02}. However, our $\kappa$ values reported in
table \ref{dse2} are within the figures discussed in 
\cite{pawlowski,fischer04}.

\begin{table}[h]
\begin{tabular}{lccccc}
\hline
    \tablehead{1}{c}{b}{Lattice}
  & \tablehead{1}{c}{b}{$\kappa$}
  & \tablehead{1}{c}{b}{$\Lambda_{QCD}$ (MeV)}
  & \tablehead{1}{c}{b}{$a_1$}
  & \tablehead{1}{c}{b}{$a_2$} 
  & \tablehead{1}{c}{b}{$\chi^2 /d.o.f.$}  \\
\hline
$16^3\times 128$ & $0.5439^{+36}_{-41}$ &  $352^{+4}_{-4}$ & $0.0063^{+4}_{-3}$ & $2.43^{+2}_{-1}$  & $1.74$ \\
$16^3\times256$ & $0.5314^{+25}_{-24}$ & $354^{+3}_{-3}$  & $0.0065^{+4}_{-3}$ & $2.40^{+1}_{-2}$ & $1.56$ \\
\hline
\end{tabular}
\label{dse2}
\caption{Fits to all lattice data.}
\end{table}

We are currently involved in improving the statistics of our analysis and, 
simultaneously, trying to understand the role of Gribov copies\footnote{In
\cite{gribovgluon}, the effect of the Gribov copies was estimated as a 2 to 
$3\sigma$ effect. This does not change our predictions for the zero momenta gluon propagator.} 
relying on the method discussed in \cite{ceasd}. We plan also to compute the 
ghost propagator and running coupling constant directly from the lattice.

%%%%%%%%%%%%%%%%%%%%%%%%%%%%%%%%%%%%%%%%%%%%%%%%
%% BACKMATTER
%%%%%%%%%%%%%%%%%%%%%%%%%%%%%%%%%%%%%%%%%%%%%%%%

\vspace*{0.2cm}  %27

%\begin{theacknowledgments}
%We thank the organizers for the nice conference in a beautiful place.
P.J.Silva acknowledges financial support from FCT via grant SFRH/BD/10740/2002.
%\end{theacknowledgments}

%\vspace*{-0.3cm}

\end{document}